\documentclass[twocolumn,trackchanges]{aastex701}

\usepackage{amsmath}
\usepackage{multirow}
\usepackage{graphicx}
\usepackage{booktabs}
\usepackage{color}

\newcommand{\CaII}{{\ion{Ca}{2}}}
\newcommand{\halpha}{H$\alpha$}
\newcommand{\hbeta}{H$\beta$}
\newcommand{\hgamma}{H$\gamma$}

\newcommand{\mdot}{$\dot{\text{M}}$}

\newcommand{\ri}{R$_{\rm i}$}

\newcommand{\Dr}{$\Delta \rm r$}
\newcommand{\tmax}{T$_{\rm max}$}

\newcommand{\msunyr}{M_{\sun} \, \rm{ yr^{-1}}}

\newcommand{\teff}{T$_{\rm eff}$}

\usepackage[export]{adjustbox}
\newcommand{\arepaemoji}{
{\includegraphics[width=12pt, valign=m]{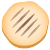}}
}

\begin{document}

\title{AREPAS\arepaemoji: A Resource for Exploring Protostellar Accretion Systems
 - Data Release I}

\author[0000-0001-8022-4378]{Marbely Micolta}
\affiliation{Department of Astronomy, University of Michigan, 1085 South University Avenue, Ann Arbor, MI 48109, USA}
\email[show]{micoltam@umich.edu}

\author[0000-0003-4507-1710]{Thanawuth Thanathibodee}
\affiliation{Department of Physics, Faculty of Science, Chulalongkorn University, 254 Phayathai Road, Pathumwan, Bangkok, 10330 Thailand.}
\email[]{thanawuth.t@chula.ac.th}

\author[0000-0003-2294-4187]{Katya Gozman}
\affiliation{Department of Astronomy, University of Michigan, 1085 South University Avenue, Ann Arbor, MI 48109, USA}
\email[]{kgozman@umich.edu}

\author[0000-0002-3950-5386]{Nuria Calvet}
\affiliation{Department of Astronomy, University of Michigan, 1085 South University Avenue, Ann Arbor, MI 48109, USA}
\email[]{ncalvet@umich.edu}

\submitjournal{Research Notes of the AAS}

\begin{abstract}
We present AREPAS\arepaemoji, a visualization tool for the exploration of the first data release of the
open library of magnetospheric accretion models for T Tauri Stars. The dataset covers the typical observed range of spectral types, mass accretion rates and inclinations of typical protoplanetary disks. This data release includes the emission lines: \halpha, \hbeta, \hgamma, Pa$\beta$, Pa$\gamma$, Pa$\delta$, Br$\gamma$, \CaII\ K, \CaII\ 8498 {\AA}, \CaII\ 8542 {\AA}. AREPAS\arepaemoji allows for parameter exploration and the comparison of models to user-input observations. 
\end{abstract}

\received{March 10, 2026}
\accepted{March 11, 2026}

\section{Introduction} 

Stars are born from the collapse of molecular clouds,
surrounded by protoplanetary disks which are the natural by-products of star formation due to angular momentum conservation. (Classical) T Tauri Stars, or (C)TTS, are young low-mass stars that accrete mass from their disk, shaping the early evolution of both the star and disk and by proxy the formation of planets as well.

Magnetospheric accretion is the prevailing model for the accretion of material from the protoplanetary disk onto the TTS. In this paradigm, the magnetic field of the star truncates the inner disk and matter is guided by the field lines at free fall velocities, until it impacts the photosphere in an accretion shock \citep[see][for a review]{hartmann_accretion_2016}. While the emission from the shock is visible as a UV, 
and in some cases optical, excess over the stellar photosphere \citep[][]{calvet_structure_1998}, strong profiles with broad wings and occasional red-shifted absorption in the emission lines is one of the most characteristic features of accreting stars. These lines form in the accretion flows and play an important role in measuring mass accretion rates 
\citep[][]{muzerolle_emissionline_2001, white2003,natta2004,thanathibodee_complex_2019,thanathibodee_censusII_2023} and in characterizing the composition of the gas accreting onto the star \citep[][]{micolta_ca_2023,micolta_using_2024,thanathibodee_model_2024,micolta_vanishing_2025}.

To better understand this process across different star-forming regions (ages), stellar parameters, accretion parameters, and different emission lines, we have produced an open library containing a large grid of magnetospheric accretion models 
\citep[see \S\ref{sec:mod};][]{micolta_database_2026}, 
Additionally, we developed a visualization tool, \href{https://arepas.streamlit.app/}{AREPAS}\arepaemoji, allowing the user to explore the model grid parameter space in an interactive and accessible way, as well as to directly compare between models and user-input observations (see \S \ref{sec:access}). This tool will enable a broad range of transformative research.

\section{Magnetospheric accretion models} \label{sec:mod} 


To calculate the structure and emission of the magnetospheric accretion flows, we used 
{\it CV-multi} \citep{hartmann_magnetospheric_1994,muzerolle_br_1998,muzerolle_emissionline_2001}. In this framework, the accretion flows have a dipolar, axisymmetric geometry characterized by the inner radius (\ri) and the width at the base of the flow (\Dr) in the disk. The density of the flow is determined by the geometry and the mass accretion rate (\mdot). The temperature distribution of the flow is semi-empirical, obtained from balancing a volumetric heating rate ($r^{-3}$) and an optically thin cooling law \citep[][]{hartmann_magnetospheric_1994}, parametrized by the maximum flow temperature (\tmax) and constrained by the accretion rate \citep[][]{muzerolle_emissionline_2001}. The mean intensities are calculated using the extended Sobolev approximation, which are used to calculate the radiative rates in the statistical equilibrium equations. The line profiles are calculated using the ray-by-ray approach for a given inclination angle $i$ from the line of sight, and assuming Voigt profiles.

In our first data release (DR1), we are including models for hydrogen and calcium II lines. To calculate populations and optical depths, the models adopt a 16-level hydrogen atom \citep[][]{muzerolle_emissionline_2001} and a 5-level calcium atom. Additionally, for Ca, its abundance relative to H in the flows is left as a free parameter \citep{micolta_ca_2023,micolta_using_2024},
and we cite it relative to the solar value \citep[$\rm log \left(N_{Ca} / N_{H}\right)_{\odot} = 6.31 \pm 0.04$,][]{asplund_cosmic_2005}.


\begin{deluxetable*}{l l l l |l l l l}[t!]
\centering
\tablecaption{Parameter space of the models \label{tab:model_param}}
\tablehead{
\multicolumn{4}{c}{Stellar Parameters} & \multicolumn{4}{c}{Magnetospheric accretion parameters} 
}
\startdata
SpT & \teff (K) & R (R$_\odot$) & M (M$_\odot$) & Parameter & Min & Max & Step \\
\hline
K2 & 4900 & 1.92 & 1.43 & log {\mdot} ($\rm \msunyr$) & -10.00 & -7.00 & 0.25 \\
K5 & 4350 & 1.52 & 0.87 & T$_{\rm max}$ (K)	 & 6500 & 14000 & 500 \\
K7 & 4060 & 1.45 & 0.73 & R$_{\rm i}$	(R$_{\star}$) & 2.0 & 6.0 & 0.5 \\
M1 & 3705 & 1.43 & 0.61 & $\Delta\rm r$ (R$_{\star}$) & 0.5 & 2.0 & 0.5 \\
M3 & 3415 & 1.33 & 0.47 & $i$ (deg)	 & 15 & 75 & 15 \\
M5 & 3125 & 1.17 & 0.31 & $[$Ca/H$]^*$ & -2.0 & 0.0 & ... \\
\hline
\enddata
\tablenotetext{*}{$\rm [Ca/H] = log (N_{Ca}/N_{H}) - log (N_{Ca}/N_{H})_{\sun}$, where N refers to the abundances by number. The available range for \CaII\ lines is: $\rm [Ca/H] = 0.0, -0.3, -1.0, -2.0$, which corresponds to abundances 100\%, 50\%, 10\% and 1\% solar values}
\end{deluxetable*}

\section{DATA RELEASE I} \label{sec:access}

The DR1 model grid covers six spectral types, spanning the typical range for CTTS of nearby star forming regions \citep[e.g.][]{manara_demographics_2023}. The stellar parameters were estimated based on a 3 Myr PARSEC evolutionary model isochrone \citep[][see Table \ref{tab:model_param}]{bressan_parsec_2012}. For each model star, our database includes the lines \halpha, \hbeta, \hgamma, Pa$\beta$, Pa$\gamma$, Pa$\delta$, Br$\gamma$, \CaII\ K, \CaII\ 8498 {\AA}, \CaII\ 8542 {\AA}.
For each emission line, we cover the full range of magnetospheric accretion parameters listed in Table \ref{tab:model_param}. 
The raw dataset is publicly available\footnote{ \href{https://github.com/marmicolta/database_magneto_models}{Dataset Repository }\citep{micolta_database_2026}}. 

We also developed the visualization tool \href{https://arepas.streamlit.app/}{AREPAS}\arepaemoji in order to make the exploration of the full dataset and parameter space accessible. 
This app is publicly available through any web browser and is built using Python in the Streamlit framework. It is open-source and the code can be found in this Github repository\footnote{\url{https://github.com/marmicolta/AREPAS}}. Users can select any combination of specific lines, accretion rates, maximum flow temperatures, disk inner radii, disk widths, inclinations, and spectral types and plot the specific line profiles on an interactive graph with zoom and hover capabilities. Users can select and deselect as many profiles to plot as they wish, and they can either plot profile flux on the stellar (model) surface (in erg/cm$^2$/s/{\AA}), the flux normalized to the continuum, or the luminosity above the continuum (in erg/cm$^2$).

There is also an option for users to upload their own files that contain e.g. an observed line profile and compare it to the model profiles. This file must be in a tabular format such that the first three columns are velocity (in km/s),  flux (in erg/cm$^2$/s/{\AA}) and distance (in pc). The third column only needs to contain the distance value in the first cell, other cells may be filled in but they will be ignored.
Likewise, the file can contain other columns, but anything past the first three columns will be ignored. Because model fluxes are calculated at the surface of the star, when comparing observational and model line profiles, it is important to compare using \textit{luminosities} or the \textit{normalized} profiles --- the app will automatically convert from input flux to either one selected. 
Observations need to be corrected by reddening before being uploaded.

Users can also easily save the raw data of any subset of profiles that are selected as a csv file for further analysis or $\chi^2$ fitting. This app is not meant for rigorous fitting of line profiles, but rather intended for users to build intuition on how profiles change with varying parameters and to estimate a coarse grid of parameters that may best be used as a starting point to fit an observed profile. Any comments, bug finds, or suggestions for improvement of the app itself are welcome and can be reported as a \href{https://github.com/marmicolta/AREPAS/issues}{Github issue} in the app repository.

\begin{acknowledgments}
This work was partially supported by NASA/FINESST 80NSSC24K1550, HST-AR-17047.001-A, and NASA XRP 80NSSC2K0151 grants.
\end{acknowledgments}

\bibliography{references}{}
\bibliographystyle{aasjournalv7}

\end{document}